\documentclass[aps,twocolumn]{revtex4}

\begin{document}

\title{Comment on ``Can gravity distinguish between Dirac and
Majorana neutrinos?''}

\author{Jos\'e F. Nieves$^1$ and Palash B. Pal$^2$\\ 
$^1$Department of Physics, P.O. Box 23343, 
University of Puerto Rico, R\'{\i}o Piedras,
Puerto Rico 00931-3343\\ 
$^2$Saha Institute of Nuclear Physics, 
1/AF Bidhan-Nagar, Calcutta 700064, India}
\affiliation{}
\date{October 2006}

\maketitle

In a recent letter\cite{singh}, Singh, Mobed and Papini (hereafter
referred to as (I)) claim to show that the gravitaional field
associated with the Lense-Thirring metric can distinguish between
Dirac and Majorana neutrinos.  According to these authors, the matrix
element of the interaction Hamiltonian is significantly different,
depending on the Dirac or Majorana nature of the neutrino.

Here we point out that the treatment of Majorana neutrinos in (I) is
not valid, due to an incorrect definition and construction of the
spinor that corresponds to a Majorana particle.  As a consequence, the
results of this paper concerning Majorana neutrinos are not valid, and
in particular the major claim stated above does not follow.

The Majorana condition is a requirement on the fermion field operator
$\psi$, viz., that it should be self-conjugate up to a phase factor
$\xi$,
\begin{equation}
\label{majcond}
\psi^c(x) = \xi \psi(x)\,,
\end{equation}
where the superscript $c$ indicates the operation of the
Lorentz-invariant complex conjugation. Instead of Eq.\
(\ref{majcond}), the authors of (I) imposed the self-conjugacy
condition on the spinors to construct a spinor $W$ for the Majorana
case. This is incorrect, and in particular the spinor $W$ is not a
solution of the free particle Dirac equation.

To be more specific, let us consider their Eqs.\ 10 and 11.  In a more
conventional notation, the authors write the spinor for the Dirac case
in the form
\begin{equation}
u(k) = u_L + u_R \,,
\end{equation}
where $u_{L,R}$ are the left and right chiral projections of $u(k)$.
These two components satisfy
\begin{eqnarray}
\label{diraceq}
\rlap/k u_L = m u_R\,,\qquad
\rlap/k u_R = m u_L\,,
\end{eqnarray}
so that $u(k)$ satisfies the Dirac equation $\rlap/k u = mu$.  The
spinors chosen in (I) for the Majorana case are given in their Eq.\ 11,
which in our notation is
\begin{eqnarray}
\label{W}
W_{1,2} & = & u_L \pm (u_L)^c\,.
\end{eqnarray}
This construction is meaningless.  In fact, using Eq.\ (\ref{diraceq})
it can be verified that the spinors $W_{1,2}$ are not solutions of the
free particle equation $\rlap/k W_{1,2} = m W_{1,2}$, as they should
be.

A Majorana free particle spinor satisfies the same equation as the
Dirac free particle spinor but, in accord with Eq.\ (\ref{majcond}),
the four linearly independent solutions appear in the plane wave
expansion of the Majorana field in the form
\begin{equation}
\label{planewaveexp}
\psi(x) = \int \! {d^3p \over N_p}
\Big( a_\lambda(p) u_\lambda(p) e^{-ip\cdot x} + \xi^*
a^\dagger_\lambda(p) 
v_\lambda(p) e^{ip\cdot x} \Big), 
\end{equation}
with an implied sum over helicities $\lambda$, $v = u^c$, and the
factor $N_p$ that depends on the normalization of the spinors.  For a
Dirac field, the creation operator in Eq.\ (\ref{planewaveexp}) would
be $b^\dagger_\lambda$, which is not the hermitian conjugate of
$a_\lambda$, and Eq.\ (\ref{majcond}) would not hold.  For a given
operator bilinear in the Majorana field, its matrix element between
two states of momenta $p$ and $p'$ is of the form
\begin{equation}
\left< p' \left| \bar\psi O\psi \right| p \right> \propto 
\bar u(p') O u(p) - \bar v(p) O v(p') \,.
\end{equation}
Two terms are obtained since both $\psi$ and $\bar\psi$ can either
create or annihilate a particle.  This is quite a different thing than
calculating $\bar W O W$ with the spinors given in Eq.\ (\ref{W}).

Leaving behind the fact that the calculation for the Majorana case is
not correctly done in (I), there remains the question of whether or
not the relevant matrix element of the gravitational interaction term
considered in (I) is different depending on whether the neutrino is a
Dirac or a Majorana particle.  Similar questions have been posed in
the existing neutrino literature numerous times.  As is well known,
careful consideration of all the issues involved, such as the fact
that only the matrix elements between the left-handed projections are
physically relevant, shows in all cases that any differences are
proportional to factors of $m/E$, so that they vanish in the zero-mass
limit \cite{confthm}.  The situation in the present context is just
the same in this respect.  In particular, with the mean energy of the
neutrino taken to be 1 MeV in (I), any difference between the Dirac
and Majorana cases are unobservable in that context.

\end{document}